\documentclass[conference]{IEEEtran}
\usepackage[letterpaper, left=0.65in, right=0.65in, bottom=1.01in, top=0.75in]{geometry}
\IEEEoverridecommandlockouts
\usepackage[binary-units]{siunitx}
\usepackage{tabularx}
\usepackage{multirow} \usepackage{array}

\usepackage{color, colortbl}
\usepackage[binary-units]{siunitx} 
\DeclareSIUnit{\bps}{bps}
\usepackage{graphicx}
\usepackage{cite}
\usepackage{amssymb}
\usepackage{amsmath}
\usepackage[caption=false,font=footnotesize]{subfig}
\def\BibTeX{{\rm B\kern-.05em{\sc i\kern-.025em b}\kern-.08em
    T\kern-.1667em\lower.7ex\hbox{E}\kern-.125emX}}
\usepackage{flushend}

\begin{document}
%
%\title{Cell-Free Massive MIMO vs. Ultra-Massive MIMO in Terahertz Communication Networks}
\title{Cell-Free Terahertz Massive MIMO:\\ A Novel Paradigm Beyond Ultra-Massive MIMO}

% author names and affiliations
% use a multiple column layout for up to three different
% affiliations
\author{\IEEEauthorblockN{Wei Jiang\IEEEauthorrefmark{1} and Hans D. Schotten\IEEEauthorrefmark{1}\IEEEauthorrefmark{2}}
\IEEEauthorblockA{\IEEEauthorrefmark{1}German Research Center for Artificial Intelligence (DFKI)\\Trippstadter Street 122,  Kaiserslautern, 67663 Germany\\
  }
\IEEEauthorblockA{\IEEEauthorrefmark{2}Rheinland-Pf\"alzische Technische Universit\"at Kaiserslautern-Landau\\Building 11, Paul-Ehrlich Street, Kaiserslautern, 67663 Germany\\
 }
%\thanks{This work was supported by the German Federal Ministry of Education and Research (BMBF) through \emph{Open6G-Hub} (Grant no.  \emph{16KISK003K}) and \emph{AI-NET PROTECT} (Grant no.  \emph{16KIS1283}) projects.}
}

% make the title area
\maketitle
\begin{abstract}
%\boldmath
Terahertz (THz) frequencies have recently garnered considerable attention due to their potential to offer abundant spectral resources for communication, as well as distinct advantages in sensing, positioning, and imaging. Nevertheless, practical implementation encounters challenges stemming from the limited distances of signal transmission, primarily due to notable propagation, absorption, and blockage losses. To address this issue, the current strategy involves employing ultra-massive multi-input multi-output (UMMIMO) to generate high beamforming gains, thereby extending the transmission range. This paper introduces an alternative solution through the utilization of cell-free massive MIMO (CFmMIMO) architecture, wherein the closest access point is actively chosen to reduce the distance, rather than relying solely on a substantial number of antennas. We compare these two techniques through simulations and the numerical results justify that CFmMIMO is superior to UMMIMO in both spectral and energy efficiency at THz frequencies. 
 \end{abstract}

\maketitle

\section{Introduction}

\begin{figure}[!b]
\centering
\includegraphics[width=0.8\linewidth]{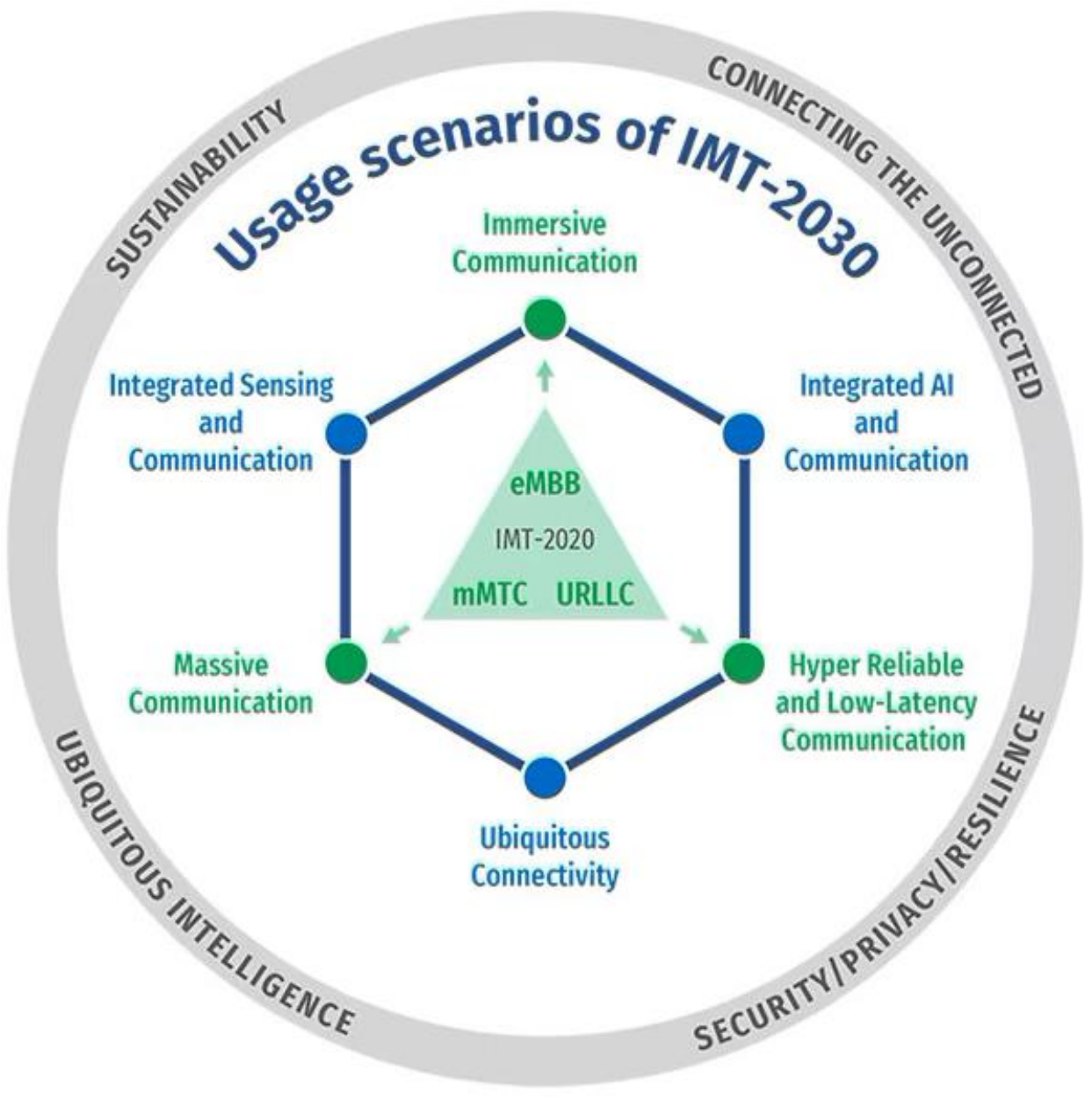}
\caption{Six usage scenarios for IMT-2030 and four overarching aspects.}
\label{Figure_6GScenarios}
\end{figure}

With an evolution dating back to the 1980s \cite{Ref_jiang2024TextBook}, the present focus in both academia and industry within cellular communications has transitioned towards driving forward the advancement of sixth-generation (6G) cellular technology, also formally referred to as IMT-2030 \cite{Ref_jiang2021road}. Notably, significant strides were made in 2023 as ITU-R Working Party 5D (WP 5D) approved the new recommendation for the IMT-2030 framework, referred to as ITU-R \textit{M.2160: Framework and overall objectives of the future development of IMT for 2030 and beyond} \cite{Ref_imt2023framework}. This framework recommendation outlines six usage scenarios for IMT-2030, as depicted in \figurename \ref{Figure_6GScenarios}, encompassing \textit{Immersive Communication}, \textit{Hyper Reliable and Low-Latency Communication}, \textit{Massive Communication}, \textit{Ubiquitous Connectivity}, \textit{Artificial Intelligence and Communication}, and \textit{Integrated Sensing and Communication}.

\renewcommand{\arraystretch}{1.2}
\begin{table*}[t]  \scriptsize 
	\centering
	\caption{Possible use of THz sensing and communications in the IMT-2030 usage scenarios \cite{Ref_jiang2024terahertz}. }
	\begin{tabular}{|l|c|c|c|c|c|c|c|c|c|c|}
		\hline
		\multirow{15}{*}{\textbf{IMT-2030 Usage Scenarios}} & 		 
		\multicolumn{7}{c|}{\textbf{THz Communications}} & \multicolumn{3}{c|}{\textbf{THz Sensing}}\\ \cline{2-11}
		 & \rotatebox[origin=c]{90}{\parbox[c][7mm][c]{30mm}{\raggedright THz Hotspot}} &  \rotatebox[origin=c]{90}{\parbox[c][7mm][c]{30mm}{\raggedright THz Campus Networks}}   & \rotatebox[origin=c]{90}{\parbox[c][7mm][c]{30mm}{\raggedright  THz D2D Link}} & \rotatebox[origin=c]{90}{\parbox[c][7mm][c]{30mm}{\raggedright THz Vehicle Networks}} & \rotatebox[origin=c]{90}{\parbox[c][7mm][c]{30mm}{\raggedright THz Security}} & \rotatebox[origin=c]{90}{\parbox[c][7mm][c]{30mm}{\raggedright THz Backhauling}} &  \rotatebox[origin=c]{90}{\parbox[c][7mm][c]{30mm}{\raggedright Nano-Networks}} & \rotatebox[origin=c]{90}{\parbox[c][7mm][c]{30mm}{\raggedright THz Sensing}} & \rotatebox[origin=c]{90}{\parbox[c][7mm][c]{30mm}{\raggedright  THz Imaging}} & \rotatebox[origin=c]{90}{\parbox[c][7mm][c]{30mm}{\raggedright  THz Localization}} \\ \hline  
   \textbf{Immersive Communication} & \checkmark  & $\circ$& \checkmark & \checkmark &  & \checkmark &  & $\circ$ &  & \checkmark\\ \hline
   \textbf{Hyper Reliable and Low-Latency Communication} &  $\circ$ & \checkmark& \checkmark & \checkmark & \checkmark& \checkmark &  & $\circ$ &  & $\circ$\\ \hline
   \textbf{Massive Communication} &   & $\circ$ &  &  &  &  & \checkmark &  &  & \checkmark\\ \hline
   \textbf{AI and Communication} &  \checkmark &\checkmark &\checkmark  &\checkmark  & \checkmark& \checkmark &  & \checkmark & \checkmark & \checkmark\\ \hline
   \textbf{Integrated Sensing and Communication} &   & &  &  & $\circ$& $\circ$ & \checkmark & \checkmark & \checkmark & \checkmark\\ \hline
   \textbf{Ubiquitous Connectivity} &   & & \checkmark &  & & \checkmark &  &  &  & \checkmark\\ \hline
		\multicolumn{11}{|l|}{\begin{tabular}[l]{@{}l@{}} Note:  \end{tabular}} \\  
		\multicolumn{11}{|l|}{\begin{tabular}[l]{@{}l@{}} The \checkmark symbol indicates that this THz application can directly contribute to the specified 6G usage scenario, the $\circ$ symbol denotes a potential contribution,\\ and a blank signifies no connection whatsoever.  \end{tabular}} \\ \hline
	\end{tabular} 
	\label{Table_THzfor6GScenarios} 
\end{table*}

The terahertz (THz) spectrum has garnered significant attention in recent years and is being investigated as a promising avenue for the development of 6G and beyond 6G systems  \cite{Ref_rappaport2019wireless}. The massive spectrum available within the THz frequencies presents prospects for ultra-high-speed wireless  communication applications. Due to the tiny wavelengths of THz signals, antennas can be miniaturized, thereby unlocking possibilities for unprecedented applications like nanoscale communications \cite{Ref_lemic2021survey}. Moreover, THz signals extend their utility beyond communication by facilitating high-definition sensing, imaging, and precise positioning within the immediate physical surroundings  \cite{Ref_sarieddeen2020next}. These unique features have sparked considerable interest recently, positioning it as a pivotal facilitator for the implementation of Integrated Sensing and Communications (ISAC) in the context of 6G and beyond. By amalgamating THz communications and THz sensing into a cohesive framework, these dual-functional wireless networks foster a synergistic relationship through the concepts of "sensing-aided communication" and "communication-aided sensing." To provide an overview of how THz communications and THz sensing can contribute to the attainment of 6G objectives, we present our perspective on the correlation between THz applications and IMT-2030 usage scenarios in \cite{Ref_jiang2024terahertz}. It is also reproduced in Table \ref{Table_THzfor6GScenarios} for easier reference.

Despite its significant potential, practical implementation faces challenges arising from limited signal transmission distances \cite{Ref_akyildiz2018combating}, primarily attributed to notable propagation, absorption, and blockage losses. To tackle this issue, the current approach involves deploying ultra-massive multi-input multi-output (UMMIMO) \cite{Ref_faisal2020ultramassive} to generate substantial beamforming gains, thereby expanding the transmission range. This paper proposes an alternative solution by employing cell-free massive MIMO (CFmMIMO) \cite{Ref_jiang2021impactcellfree}, where the closest access point (AP) is actively selected to reduce the distance, rather than relying solely on a large number of antennas. Through simulations, we compare these two techniques, and the numerical results support the superiority of CFmMIMO over UMMIMO in both spectral and energy efficiency at THz frequencies.

The paper is structured as follows: The subsequent section will clarify the primary issues limiting the transmission range of THz signals, with Section III introducing UMMIMO. The proposed CFmMIMO-THz paradigm will be presented in Section IV, followed by simulations in Section V. 

\section{The Major Challenges}

Although the substantial potential of THz communications and sensing in the context of 6G and beyond is recognized, the practical implementation faces challenges arising from limited signal transmission distances, primarily associated with the following issues:

\subsubsection{High Spreading Loss}
When an isotropic radiator emits an electromagnetic wave, energy uniformly spreads across a sphere's surface. The \textit{Effective Isotropic Radiated Power (EIRP)} indicates maximum energy emitted in a specific direction compared to a theoretical isotropic antenna. It results from multiplying transmission power by the transmitting antenna's gain in the receiving direction. As a result, extremely high path loss at THz frequencies rises from the tiny aperture area of the receive antenna, which is proportional to the wavelength of its operating frequency \cite{Ref_jiang20226GCH5}.  

\subsubsection{Atmospheric Absorption}
\begin{figure}[!tbph]
\centering
\includegraphics[width=0.95\linewidth]{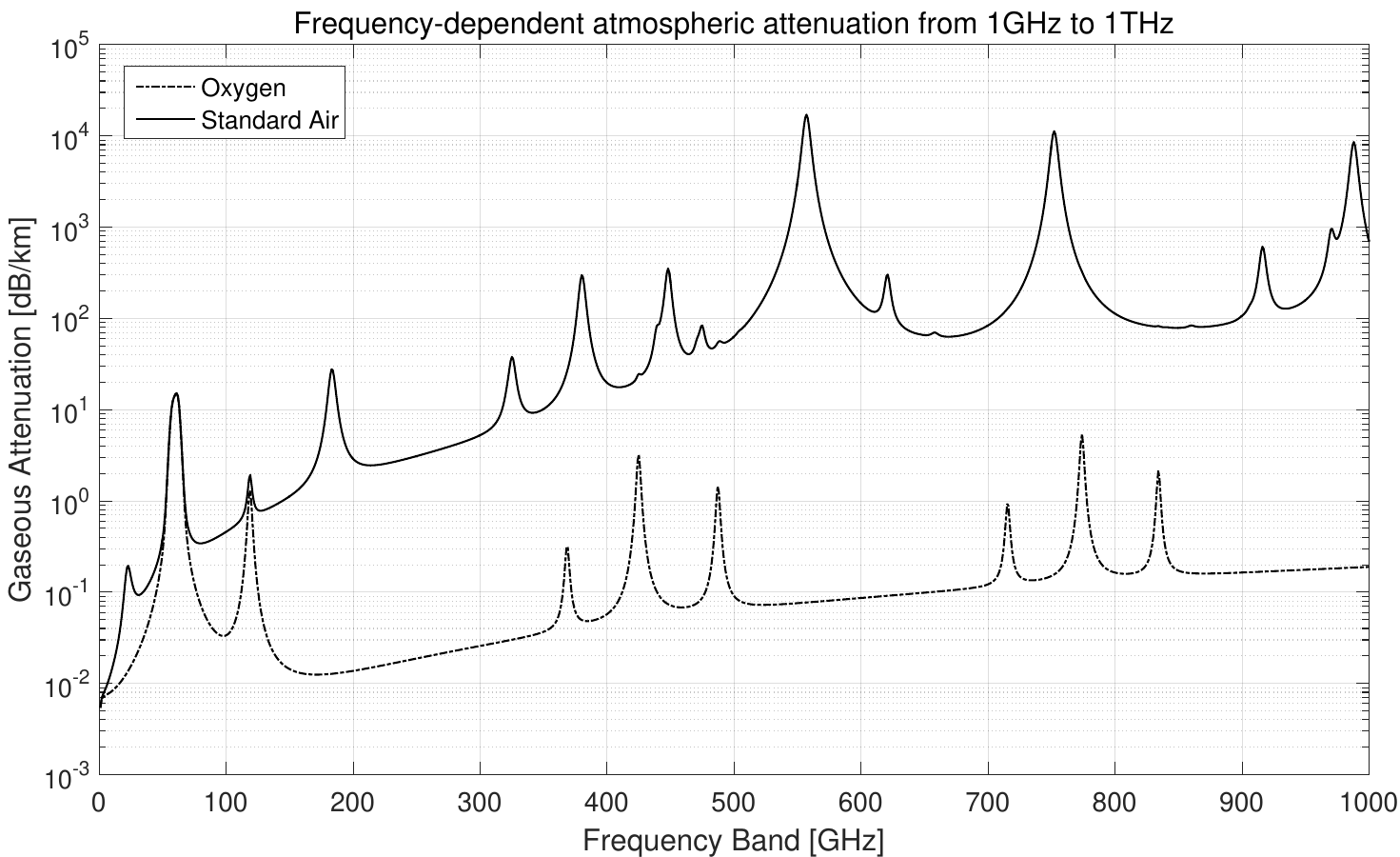}
\caption{Illustration of atmospheric absorption \cite{Ref_jiang2023terahertz}. } 
\label{Diagram_atmosphericAbsorption}
\end{figure}

Although gaseous molecules do absorb some electromagnetic wave energy, the impact of atmospheric absorption on frequencies below 6GHz is minimal, and conventional cellular systems generally do not consider it in link budget calculations. However, this phenomenon becomes notably pronounced with THz waves, where absorption losses sharply increase at specific frequencies. This attenuation arises from the interaction between electromagnetic waves and gaseous molecules. As the wavelength of THz waves approaches the size of atmospheric molecules, the incoming wave induces rotational and vibrational transitions within polar molecules. These processes exhibit quantum characteristics, displaying resonant behaviors at distinct frequencies based on the molecular structure. Consequently, there are significant absorption peaks at specific frequencies \cite{Ref_slocum2013atmospheric}. Oxygen, a primary component of the atmosphere, plays a crucial role in atmospheric absorption, especially in clear air conditions. Simultaneously, the presence of water vapor in the air strongly influences the propagation of electromagnetic waves. Fig. \ref{Diagram_atmosphericAbsorption} illustrates the atmospheric attenuation from \SIrange{1}{1000}{\giga\hertz} \cite{Ref_WJ_itu2019attenuation}.

\subsubsection{Weather Effects}

In addition to gaseous absorption, another atmospheric factor to account for in outdoor environments is the impact of weather. Similar to the effect of water vapor, research indicates that suspended liquid water droplets, present in various forms like clouds, fog, snowflakes, or falling rain, can absorb or scatter incident signals. This occurs because their physical dimensions align with the scale of THz wavelengths. Clouds consist of tiny water particles, some as small as \SI{1}{\micro\meter}, or ice crystals ranging from \SIrange{0.1}{1}{\milli\meter} in size. Water droplets in raindrops, fogs, hailstones, and snowflakes typically take the form of oblate spheroids with radii reaching a few tens of millimeters, or nearly perfect spheres with radii below 1 \SI{1}{\milli\meter}. Given that the dimensions of these water droplets correspond to THz wavelengths (ranging from \SIrange{0.1}{1}{\milli\meter}), they effectively diminish the power of THz waves through absorption and scattering. While this attenuation may not be as prominent as path loss or atmospheric absorption, it remains a consideration for system design \cite{Ref_weng2021millimeter}.

\subsubsection{Line-of-Sight Blockage}
Due to its short wavelengths in the THz range, surrounding physical objects attain sufficiently large dimensions for scattering effects, making it challenging to achieve specular reflections on typical surfaces. In contrast, THz systems heavily depend on narrow pencil beams to amplify the effective propagation distance. As a result, a line-of-sight (LoS) path between the transmitting and receiving ends is highly preferable. However, LoS THz links are considerably more susceptible to obstruction from both macro-scale elements like buildings, furniture, vehicles, and foliage, as well as micro-scale objects, including humans, when compared to the conventional sub-6GHz band \cite{Ref_rappaport2014millimeter}. Consequently, it becomes essential to thoroughly comprehend the characteristics of blockages and formulate effective strategies to prevent them or swiftly restore connectivity in the event of an obstruction.

\section{Ultra-Massive MIMO}
Apart from experiencing significant propagation, atmospheric, and blockage losses, the challenge of limited transmission range is exacerbated by the following two factors \cite{Ref_peng2020channel}:
\begin{itemize}
\item \textit{Strong thermal noise}: The power of additive white noise is directly related to the signal bandwidth, maintaining a constant power density. Consequently, the distinctive benefit of extremely large bandwidth in THz signal transmission introduces a side effect of strong thermal noise. 
\item \textit{Hardware constraint}: The transmit power of the THz transmitter is significantly limited due to the frequency-dependent decrease in output power, anticipated to remain at decibel-milli-Watt levels in the near future. Hence, attempting to increase power for the purpose of extending communication distance is not a viable option \cite{Ref_rikkinen2020THz}.  
\end{itemize}

In order to surpass the signal transmission limitations within a few meters, the use of high-gain directional antennas \cite{Ref_yang2013random} becomes essential to offset the substantial propagation loss in THz communications and sensing. The advantage of the compact wavelengths allows for densely packing a large number of elements in a confined space, resulting in substantial beamforming gains \cite{Ref_han2021hybrid}.
The introduction of nano-antennas opens up possibilities for constructing extensive arrays in the realm of THz communications. In 2016, Akyildiz and Jornet \cite{Ref_akyildiz2016realizing} proposed the concept of UMMIMO communications, illustrating a system with a $1024\times 1024$ configuration, wherein both the transmitter and receiver employ arrays comprising 1024 nano-antennas each. However, managing a vast number of elements presents challenges, such as substantial power consumption and intricate hardware design. It is prudent to reconsider the array architecture and beamforming strategies in UMMIMO systems operating in the THz band. 

While fully digital beamforming can create desired beams, the associated energy consumption and hardware costs become impractical, as each antenna in a large-scale array requires a dedicated radio frequency (RF) chain. This prompts the exploration of analog beamforming with lower complexity. Currently, various hybrid beamforming approaches are under discussion in the literature \cite{Ref_jiang2022initial_ICC}, including a new form called an array of sub-arrays (AoSA) to strike a balance between power consumption and data rate. AoSA divides all elements into distinct subsets referred to as subarrays (SAs), with each subarray exclusively accessible to a specific RF chain \cite{Ref_jiang2022initial}. It conducts signal processing at the subarray level, utilizing fewer phase shifters, which significantly mitigates hardware costs, power consumption, and signal power loss. Moreover, the cooperative optimization of beamforming and spatial multiplexing can be achieved by integrating precoding in the baseband.

\begin{figure}[!tbph]
\centering
\includegraphics[width=0.95\linewidth]{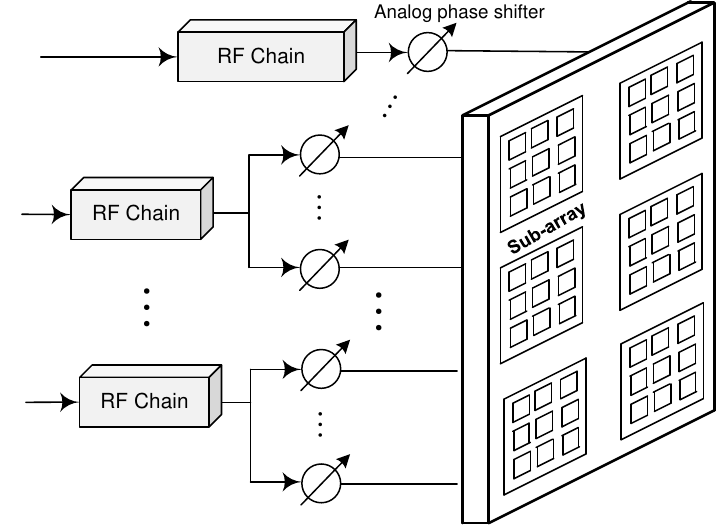}
\caption{Structure of array-of-subarrays-based UMMIMO for THz signals. } 
\label{Diagram_UMMIMO}
\end{figure}

Consider a three-dimensional UMMIMO system that integrates an extensive number of plasmonic nano-antennas within a compact footprint, utilizing nanomaterials like graphene. As shown in \figurename \ref{Diagram_UMMIMO}, the antenna array is composed of active graphene elements positioned over a shared metallic ground layer, separated by a dielectric layer. Let the array on the transmitter and receiver sides be denoted as $M_t \times N_t$ and $M_r \times N_r$ subarrays, respectively. Each subarray comprises $Q \times Q$ antenna elements. Consequently, the resulting configuration can be expressed as an $M_t N_t Q^2 \times M_r N_r Q^2$ MIMO system, achieved by vectorizing the two-dimensional antenna indices on each side.

\begin{figure*}[!tbph]
\centering
\includegraphics[width=0.9\linewidth]{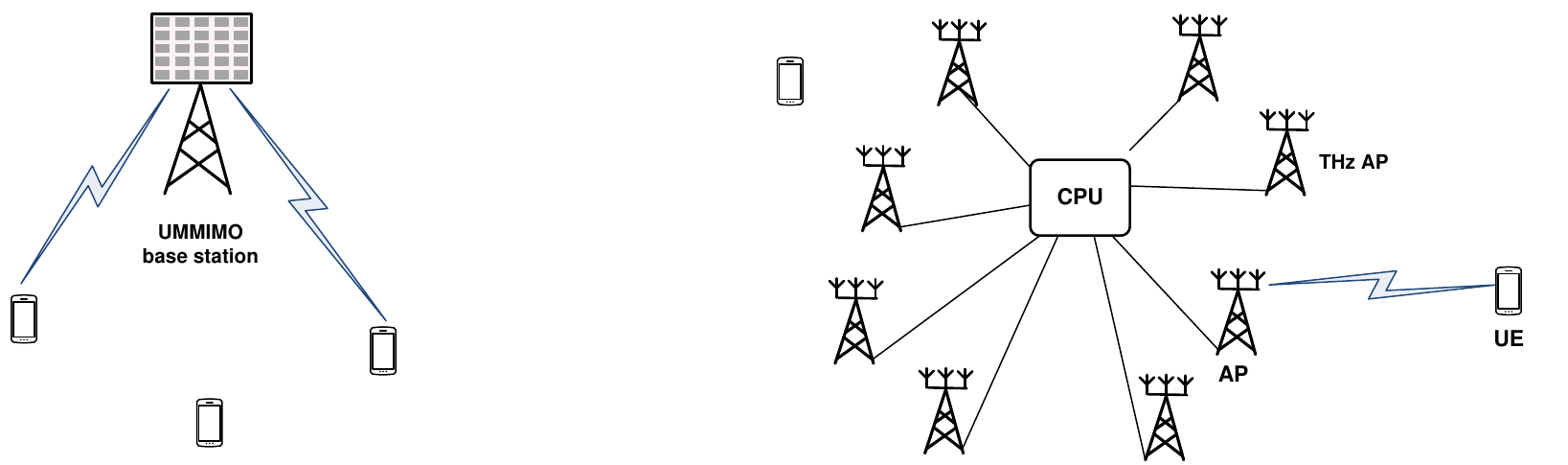}
\caption{The comparison of ultra-massive MIMO, on the left side, and cell-free Terahertz massive MIMO on the right side. The base station is equipped with a UMMIMO antenna array, suffering from long transmission distance. In contrast, the proposed CFmMIMO-THz system can substantially reduce the communication distance by selecting the closest AP for the user.  } 
\label{Diagram_cfmmmimoThz}
\end{figure*}

\section{Cell-Free Terahertz Massive MIMO (CFmMIMO-THz)}
Ngo et al. \cite{Ref_ngo2017cellfree} introduced an innovative distributed massive MIMO system characterized by a multitude of service antennas serving a small group of users. Unlike traditional cellular networks, this system lacks distinct cells or boundaries, leading to its classification as \textit{cell-free massive MIMO}. The setup initially includes $M$ access points (APs) and $K$ users across the geographical area, with a significantly larger number of APs ($M\gg K$). Each AP and user terminal is equipped with a single antenna. The APs collaborate coherently through a fronthaul network, connecting to a Central Processing Unit (CPU). This configuration enables simultaneous service to all $K$ users within the same time-frequency resource. The integration of distributed MIMO and massive MIMO concepts in this system is anticipated to leverage the inherent advantages of both technologies \cite{Ref_jiang2021cellfree}.

In contrast to UMMIMO, designed specifically for high-frequency signal transmission, CFmMIMO was crafted for low-frequency operation. Although cell-free massive MIMO systems at millimeter-wave frequencies have been studied, their adaptability to the THz band remains still missing. There is a dearth of literature on cell-free deployment over THz bands, particularly from the perspective of a massive number of antennas. Current work on the combination of THz and CFmMIMO is limited to exploring THz frequencies in backhauling for CFmMIMO \cite{Ref_abbasi2024uxnb} and cell-free spatial-spectral THz networks employing leaky-wave antennas \cite{Ref_zhu2024cellfree}. This paper introduces a novel approach by leveraging the distributed structure of a cell-free system to address the inherent limitations in the transmission distance of THz signals. 

In contrast to UMMIMO, which compensates for propagation losses through high-gain beams and a substantial number of antennas, our approach takes a proactive stance by shortening propagation distances by relocating the antenna subarrays of UMMIMO to distributed APs in a cell-free way and the selection of the closest AP for communication, as shown in \figurename \ref{Diagram_cfmmmimoThz}. With the large number of distributed APs, it is highly probable that there exist a nearby AP that offers a very short transmission distance to the user \cite{Ref_jiang2022opportunistic}. Since the large spreading and atmospheric losses of THz signals are directly related to the transmission distance, the signal quality can be substantially improved despite the scale of the antenna array at each AP is smaller than that of UMMIMO. This strategy capitalizes on the capabilities of THz frequencies, ensuring that each user is situated within a short distance from at least one AP, facilitating effective communication within the constraints of communication-range limitations. Additionally, the susceptibility of THz-band blockage can be easily mitigated in a cell-free network by implementing redundant paths.

\begin{table*}
    \centering
    \begin{tabular}{c|c|c|c|c} \hline \hline
     \multirow{2}{*}{Radius of Coverage Area [m]}    & \multicolumn{2}{c|}{Eight 8x8 SAs}    &  \multicolumn{2}{c}{Twenty 8x8 SAs} \\  \cline{2-5}
         & UMMIMO & CFmMIMO-THz  & UMMIMO & CFmMIMO-THz \\ \hline \hline
        1 &  \SI{163.3}{\giga\bps} & \SI{ 234.6   }{\giga\bps} & \SI{  176.7  }{\giga\bps} & \SI{   221.1 }{\giga\bps}\\ \hline
         2&  \SI{143.4}{\giga\bps} & \SI{ 214.8   }{\giga\bps} &  \SI{ 154.9   }{\giga\bps}& \SI{  201.6  }{\giga\bps}\\ \hline
         4&  \SI{121.7}{\giga\bps}& \SI{ 194.9   }{\giga\bps} & \SI{  134.9  }{\giga\bps} & \SI{184.4    }{\giga\bps}\\ \hline
         8&  \SI{101.6}{\giga\bps}& \SI{ 174.8   }{\giga\bps} &  \SI{  115.0  }{\giga\bps}& \SI{ 161.5    }{\giga\bps}\\ \hline
         16&  \SI{81.6}{\giga\bps}&  \SI{  154.2  }{\giga\bps}& \SI{ 94.8   }{\giga\bps} & \SI{ 141.6   }{\giga\bps}\\ \hline
         32&  \SI{61.7}{\giga\bps}& \SI{  134.8  }{\giga\bps} & \SI{ 74.8   }{\giga\bps} & \SI{ 121.6   }{\giga\bps}\\ \hline
         64&  \SI{42.1}{\giga\bps}&  \SI{ 114.8   }{\giga\bps}& \SI{ 54.9   }{\giga\bps} &\SI{ 101.6   }{\giga\bps} \\  \hline \hline
    \end{tabular}
    \caption{Numerical results in terms of achievable data rates in different coverage radii.}
    \label{tab:my_label}
\end{table*}

\section{Simulations}

The proposed CFmMIMO-THz system's performance is systematically compared with that of UMMIMO, focusing on the achievable data rate. In our simulations, we define a representative scenario where a circular area is covered by a total of eight or twenty sub-arrays, each consisting of $8\times 8$ antenna elements. The radius of the circular area varies across seven values: \SI{1}{\meter}, \SI{2}{\meter}, \SI{4}{\meter}, \SI{8}{\meter}, \SI{16}{\meter}, \SI{32}{\meter}, and \SI{64}{\meter}. This enables the observation of the performance of these two techniques at different distances.
In the traditional UMMIMO system, a base station is centrally located within the area, while the user moves randomly. In CFmMIMO-THz simulations, eight or twenty APs are distributed randomly within the coverage area, and the user is also placed randomly. The user equipment is assumed to be equipped with a single $8\times 8$ antenna array. During each simulation epoch, the locations of APs and users undergo random changes. The user consistently selects the closest AP as the access point, while the other APs can be turned off to save energy. This implies that CFmMIMO-THz demonstrates greater efficiency than UMMIMO in terms of energy consumption.

The operational frequency is configured at \SI{300}{\giga\hertz}, with a signal bandwidth of \SI{10}{\giga\hertz}. The white noise power density is $-174\mathrm{dBm/Hz}$, and the transmit power is set to $5\mathrm{dBm}$. Both the transmit and receive antenna gain are specified as $20\mathrm{dBi}$. The molecular absorption coefficient, serving as a unique THz signature, is expressed in terms of the system temperature in Kelvin, the reference temperature (296.0 Kelvin), the temperature at standard pressure (273.15 Kelvin), the system pressure in atmospheres, the reference pressure, the Planck constant, the Boltzmann constant, the gas constant, and the Avogadro constant. This simulation employs the molecular absorption information provided by the HITRAN-based model. For additional details regarding the setup of THz signal transmission, refer to the TeraMIMO simulator introduced by S. Tarboush et al. \cite{Ref_tarboush2021teramimo}.

\begin{figure}
        \centering
        \includegraphics[width=0.99\linewidth]{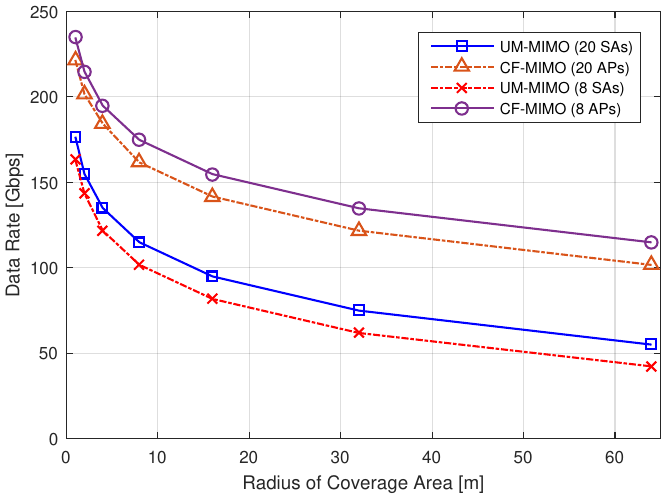}
        \caption{Performance comparison between UMMIMO and CFmMIMO-THz in terms of data rates as a function of communication distances.}
        \label{fig:enter-label}
    \end{figure}

The results depicted in \figurename \ref{fig:enter-label} and elaborated in Table \ref{tab:my_label} provide a comprehensive examination of the data rates achieved by UMMIMO and CFmMIMO-THz across diverse scenarios. Specifically, UMMIMO, employing a configuration of 8 SAs with 64 elements on each SA, attains a data rate of \SI{163.3}{\giga\bps} when the coverage radius is \SI{1}{\meter}. In stark contrast, the CFmMIMO-THz system, leveraging 8 distributed APs with 64 elements on each AP, surpasses UMMIMO's performance by a significant margin, reaching an outstanding data rate of \SI{234.6}{\giga\bps}. A noteworthy aspect of CFmMIMO-THz's superiority lies in its shortened transmission distance, where only one AP, positioned in proximity to the user, transmits THz signals. The remaining 7 APs in CFmMIMO-THz remain turned off their transceivers, thereby enhancing overall energy efficiency.
As the coverage area expands, it is anticipated that the achievable data rates exhibit a natural decline. However, CFmMIMO-THz consistently outshines UMMIMO across all observed distances. For instance, at a coverage radius of \SI{64}{\meter}, the performance metrics reveal that UMMIMO and CFmMIMO-THz deliver data rates of \SI{42.1}{\giga\bps} and \SI{114.8}{\giga\bps}, respectively. This clear performance disparity underscores the remarkable advantages of CFmMIMO-THz, showcasing its prowess in maintaining superior data transmission capabilities while improving energy efficiency.

As the antenna array scale undergoes an expansion from 8 SAs to 20 SAs, the anticipated consequence is a performance boost for UMMIMO. This improvement is evident in the data rate, which escalates from \SI{42.1}{\giga\bps} to \SI{54.9}{\giga\bps} at a coverage radius of \SI{64}{\meter}. On the contrary, the performance trajectory of CFmMIMO-THz takes a distinctive turn when employing 20 APs, demonstrating a paradoxical decrease compared to its achievements with 8 APs. This outcome arises from the equal allocation of the total transmit power, set at $5\mathrm{dBm},$ among the distributed APs. With an increase in the number of APs, each AP's transmit power decreases, and coupled with a fixed 64 antenna elements per AP, the data rate experiences a reduction compared to the previous configuration. 
Despite this decrease in transmit power, an intriguing aspect emerges: the energy efficiency of CFmMIMO-THz is further enhanced. This is attributed to the fact that only one-twentieth of the total power is now consumed, compared to the previous one-eighth allocation. This efficiency improvement is noteworthy, as it aligns with the trend of optimizing power consumption in wireless communication technologies. Even with this trade-off in power allocation, the data rates achieved by CFmMIMO-THz remain substantially superior to those of UMMIMO. For instance, at a coverage radius of \SI{64}{\meter}, CFmMIMO-THz attains a commendable data rate of \SI{101.6}{\giga\bps}, surpassing UMMIMO's \SI{54.9}{\giga\bps}.

\section{Conclusions}
This paper focuses on a pivotal challenge in THz communication and sensing, stemming from constraints in signal transmission distances. In contrast to the prevalent approach of employing ultra-massive multi-input multi-output (UMMIMO) for substantial beamforming gains to extend the transmission range, we introduce an innovative solution involving the use of a distributed massive MIMO architecture as cell-free massive MIMO (CFmMIMO). This choice is motivated by the close relationship between extreme propagation and absorption losses with propagation distance. Our proposed approach involves redistributing the substantial number of antennas across distributed APs and actively selecting the closest AP to minimize distances, moving away from dependence solely on a high number of antennas. Through simulations, we conduct a comparative analysis of these two techniques, highlighting that CFmMIMO outperforms UMMIMO in terms of both achievable data rates and energy consumption.

\bibliographystyle{IEEEtran}
\bibliography{IEEEabrv,Ref_MeditCom24.bib}

\end{document}